\documentclass[review]{elsarticle}

\usepackage{lineno,hyperref}
\modulolinenumbers[5]
\usepackage{graphicx,color}
\usepackage{wrapfig}
\usepackage{dcolumn}
\usepackage{gensymb}

\journal{ArXiv}

\begin{document}

\begin{frontmatter}

\title{Experimental Observation of a Magnetic Interfacial Effect}

\author{Wibson W. G. Silva, Sérgio V. B. Degiorgi and José Holanda$^{ }$\corref{mycorrespondingauthor}}
\cortext[mycorrespondingauthor]{Corresponding author: joseholanda.silvajunior@ufrpe.br}
\address{Programa de Pós-Graduação em Engenharia Física, Universidade Federal Rural de Pernambuco, 54518-430, Cabo de Santo Agostinho, Pernambuco, Brazil.}

\begin{abstract}
We observed a magnetic interfacial effect due to the coupling between two interfaces of different materials. The interface is compoust of an antiferromagnetic and other quasi-ferromagnetic material. This effect we measured through the ferromagnetic resonance technique without and with electric current. 
\end{abstract}

\begin{keyword}
\texttt{}Magnetic\sep non-collinear\sep uncompensated\sep interface

\end{keyword}

\end{frontmatter}
\vspace{1cm}
{\fontsize{20}{\baselineskip} \textbf{A}}ntiferromagnetic materials exhibiting topologically protected states in their momentum-space bands as well as those exhibiting topologically non-trivial real-space at the interface, inspire the production of new devices in antiferromagnetic spintronics [1-6]. A non-collinear antiferromagnet that has a giant magnetocrystalline anisotropy energy of 10 meV per unit cell is the IrMn$_3$ [7-12]. The IrMn$_3$ has a giant spin Hall angle (up to $\approx$ 0.35) in the direction [001] and crystallizes in a face-centered cubic lattice [10]. The local ferromagnetism from uncompensated spins can induce antiferromagnetic symmetry in the adjacent material [4, 13-15]. Platinum (Pt) can be an excellent adjacent material for this coupling because Pt is quasi-ferromagnetic according to Stoner's criterion [16]. Thus, IrMn$_3$ and Pt are promissory candidates for the detection of resonance signals due to the ferromagnetism at the interface. The detection of these signals can be performed using the ferromagnetic resonance (FMR) technique without and with electric current [17-29]. In this Letter, we observe ferromagnetic resonance signals at the sample of IrMn$_3$/Pt bilayer without and with electric current. Furthermore, we analyzed the change of the interface damping as a function of the electric current.

All IrMn$_3$ films were epitaxially grown in a magnetron sputtering system with pressures of base approximately 1 $\times$ $10^{-7}$ torr and of argon of 3 mtorr at a temperature of 843 K on the  (100)MgO substrates. We used one Ti(2 nm) layer to protect the surface properties of IrMn$_3$ or Pt layer. We deposited the Pt layer at 373 K temperature to diffusion on IrMn$_3$ and the Ti layer at room temperature. \textbf{Fig. 1 (a)} shows an X-ray diffraction pattern for IrMn$_3$ grown on a (100)MgO substrate. We found that IrMn$_3$ has a lattice constant of (0.377 $\pm$ 0.001) nm, which is consistent with previous literature values [30]. 
\begin{figure}[h]
	\vspace{0.1mm} \hspace{0.1mm}
	\begin{center}
		\includegraphics[scale=0.41]{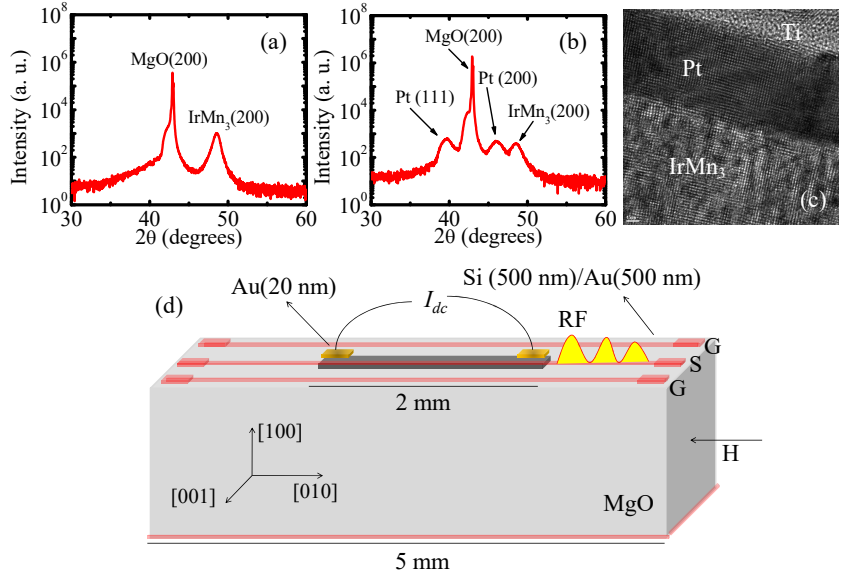}
		\caption{\label{arttype}(Color online) X-ray diffraction pattern measured for a 10 nm thick IrMn$_3$ layer without or with a 6 nm thick Pt layer capped with a 2-nm Ti layer: \textbf{(a)} IrMn$_3$(10 nm)/Ti (2 nm) and \textbf{(b)} IrMn$_3$(10 nm)/Pt(6 nm)/Ti(2 nm). \textbf{(c)} Transmission Electron Microscopy (TEM) micrograph of IrMn$_3$(10 nm)/Pt(6 nm)/Ti(2 nm). \textbf{(d)} Shows the coplanar waveguide (CPW) deposited by Electron Beam Evaporation at the up of the samples for measures of ferromagnetic resonance without and with electric current.}
		\label{puga}
	\end{center}
\end{figure}
The X-ray diffraction pattern of \textbf{Fig. 1 (b)} shows that the Pt has grown on the IrMn$_3$ surface, showing different phases. The quality of samples also were analyzed by transmission electron microscopy (TEM), as shown in the micrograph of the IrMn$_3$(10 nm)/Pt(6 nm)/Ti(2 nm) sample in \textbf{Fig. 1 (c)}. 

As it is well known, an exchange bias can be stabilized at room temperature for IrMn$_3$/Py bilayer, and thus  IrMn$_3$ is antiferromagnetically ordered at room temperature [2, 4, 10]. In the same way, the ferromagnetism in the antiferromagnetic and quasi-ferromagnetic interface is stabilized [28]. For the deposition of samples, we used a metallic mask to avoid any type of contamination on the surface of MgO, and we utilized the laser write technique for the production of the coplanar waveguide on the sample surface, as shown in the \textbf{Fig. 1 (d)}. We measured resonance signals by the flip-chip vector network analyzer ferromagnetic resonance (VNA-FMR) technique. We obtained via Lorentz fitting the frequency swept linewidths ($\Delta f_{VNA}$). Detailed steps, including the conversion from $\Delta f_{VNA}$ to the field linewidths $\Delta H$, are shown in ref. [28]. In the process of VNA-FMR, we measured the transmission coefficient by sweeping the frequency at every fixed magnetic field [4]. 

In the \textbf{Fig. 2 (a)}, we show the unit cell of IrMn$_3$, where the Mn atoms are on the \{111\} planes and their spins are aligned along the $<112>$ directions. As it is known, spins-in and spins-out configurations are nonequivalent ground states and are chiral images of each other [3, 4, 23]. For analyses of the spins-in or spins-out configurations, a transversal section can be made in the unit cell of IrMn$_3$ shown from a kagome lattice in the (111) plane with either pointing outwards in each triangular Mn arrangement (see the \textbf{Fig. 2 (b)}). This configuration represents the interface configuration of the spins-out for the film. In this case, the thin layer of Pt deposited on top of the IrMn$_3$ grows with [100] and [111] crystallographic directions (see the \textbf{Fig. 1 (b)}). 
\begin{figure}[h]
	\vspace{0.1mm} \hspace{0.1mm}
	\begin{center}
		\includegraphics[scale=0.38]{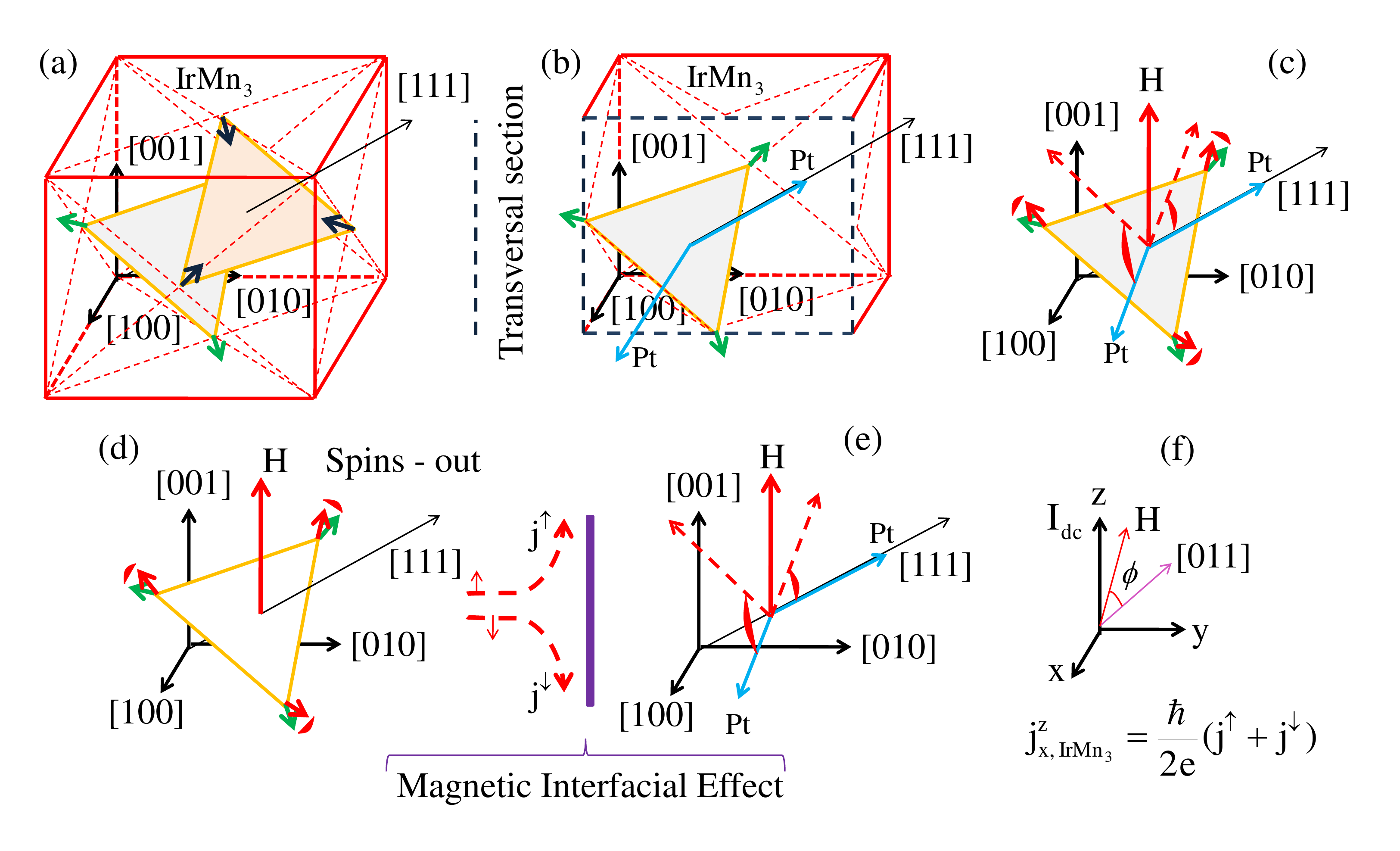}
		\caption{\label{arttype}(Color online)\textbf{(a)} Show the unit cell of IrMn$_3$ whose moments Mn are parallel to the planes \{111\} and aligned with the directions $<112>$. \textbf{(b)} The transversal section in which the interface configuration of the spins-out for IrMn$_3$. \textbf{(c)} Configuration of resonance condition of the rotation of spins of IrMn$_3$ and Pt on a magnetic field. \textbf{(d)} Arrangement from the magnetic interfacial effect of IrMn$_3$ in interface resonance condition, where a spin accumulation is created in the direction from Pt. \textbf{(e)} Arrangement from spin Hall effect of Pt in interface resonance condition, where a spin accumulation is created in the direction from IrMn$_3$. \textbf{(f)} Coordinates system of the spin currents of the antiferromagnet IrMn$_3$.}
		\label{amor}
	\end{center}
\end{figure}
The net chirality at the IrMn$_3$/Pt interface in the resonance condition can be understood by the rotation of uncompensated spins of IrMn$_3$, and the spins at the surface of Pt, as shown in \textbf{Fig. 2 (c)}. The coupling of the noncollinear spins of  IrMn$_3$, and the quasi-ferromagnetic spins of the Pt represent the ideal condition to explore the magnetic interface. [3, 4, 23, 29-35]. In resonance condition using the Hall effects of the IrMn$_3$ (see \textbf{Fig. 2 (d)}) and of the Pt (see \textbf{Fig. 2 (e)}), a spin accumulation field is created as two spin currents flowing at directions of the interface [4, 23]. The polarization of the spin accumulation field defined by the magnetic field is transversal at the magnetic interface, as shown in \textbf{Figs. 2 (d)} and \textbf{(e)}. The spin current of the IrMn$_3$ is defined as $j_{IrMn_3}^{z} = j_{y, SSS}^{z} = (j^{\uparrow} + j^{\downarrow})\hbar/2e$, which depend of the spins-in and spins-out configurations, as shows the \textbf{Fig. 2 (f)}.

The efficiency from the FMR signals is due to the net chirality of the spin structures, which provides more complexity compared to non-magnetic materials [29-36]. \textbf{Fig. 3 (a)} shows the FMR signals obtained with a VNA for one frequency of 10 GHz, magnetic field of 10.3 kOe, and different electric currents -1, 0, and +1 mA. Two observations need highlight: first, the electric current produces a significant variation in frequency swept linewidths ($\Delta f_{VNA}$), and second, a shift is caused by the accumulation field ($H_{Ac}$). In \textbf{Fig. 3 (b)} we show that the accumulation field (H$_{Ac}$) due to the electric current produces a modification in the resonance field. This is clear evidence of a magnetic interfacial effect due to the strong coupling at interface IrMn$_3$/Pt interface, similar to the IrMn$_3$/Py bilayer [4, 12, 19]. The solid curve represents the fit from the experimental data to the Kittel equation, $f = \gamma[(H_{R})(H_{R}+4 \pi M_{eff} \pm H_{Ac})]^{1/2}$, where the gyromagnetic ratio is $(\gamma_{IrMn_3/Pt})$/$\gamma_{IrMn_3/Py}$ = $5\%$, the spectroscopic splitting factor calculated considering the Stoner`s criterion [16] is $g_{IrMn_3/Pt}$/$g_{IrMn_3/Py}$ = $5\%$, $\mu_{B}$ is the Bohr magneton, $\hbar$ is the reduced Planck constant, and $4 \pi M_{eff} = 4 \pi M_S + H_{AS}$ is the effective magnetization that is much larger than the saturation magnetization $4 \pi M_S$ due to the effect of the surface anisotropy field $H_{AS}$. Using the fit with the Kittel equation to zero electric currents, we obtained for effective magnetization $4 \pi M_{eff} = (523.78 \pm 0.005)$ kOe, which is smaller than that obtained for $IrMn_3/Py$ bilayer [4, 10, 12, 19].

\begin{figure}[h]
	\vspace{0.1mm} \hspace{0.1mm}
	\begin{center}
		\includegraphics[scale=0.34]{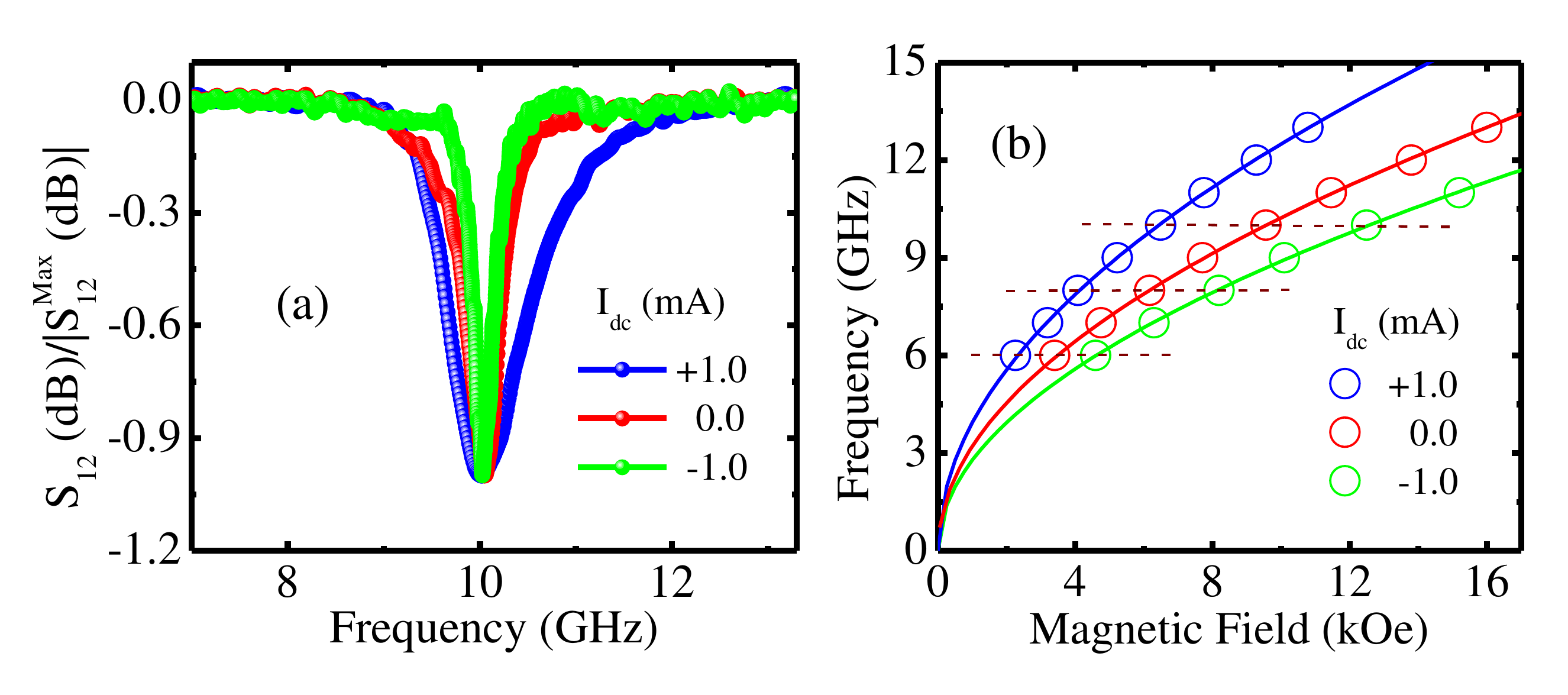}
		\caption{\label{arttype}(Color online) \textbf{(a)} Ferromagnetic resonance (FMR) signals were obtained using a VNA for one frequency of 10 GHz, magnetic field of 10.3 kOe, and different electric currents -1, 0, and +1 mA. \textbf{(b)} FMR frequency as a function of the magnetic field, which the accumulation field (H$_{Ac}$) due to the electric current produces a change in the resonance field. The fits are performed with the Kittel equation, where $(\gamma_{IrMn_3/Pt})$/$\gamma_{IrMn_3/Py}$ = $5\%$ and $4 \pi M_{eff} = (523.78 \pm 0.005)$ kOe.}
		\label{paz}
	\end{center}
\end{figure}

The properties of noncollinear antiferromagnetic materials with magnetic topological states yield large changes [31-35], this also is a characteristic of uncompensated spins that induces magnetism. In the \textbf{Fig. 4 (a)}, we represent the change from resonance frequency as a function of the spin accumulation field ($H_{Ac}$) for electric current $\pm$1 mA. The spin accumulation field increase by obeying a Kittel equation describes by greens lines. In this case, the spins-in and spins-out configurations exhibit the same energies, and both exist spontaneously in the material [31]. On the other hand, in the \textbf{Fig. 4 (b)} we show the spin accumulation field as a function of electric current in the resonance frequency of 10 GHz. A small onset of saturation of the spin accumulation field is observed for electric currents $I_{dc} <$ -1mA and $I_{dc} >$ + 1mA, which is also observed in other bilayers [4, 10, 11, 27]. The polarization from the spin accumulation field obeys the electric current confirming the process from the manipulation of the magnetic interfacial effect in the IrMn$_3$/Pt bilayer, similar to the manipulation of the exchange bias at the antiferromagnetic/ferromagnetic bilayer [4, 12, 19]. In the \textbf{Fig. 4 (c)}, the fit is made with the expression $\Delta H = (\alpha / \gamma)f$ [33], where $\alpha$ is the magnetic Gilbert damping of the interface. 

\begin{figure}[h]
	\vspace{0.1mm} \hspace{0.1mm}
	\begin{center}
		\includegraphics[scale=0.34]{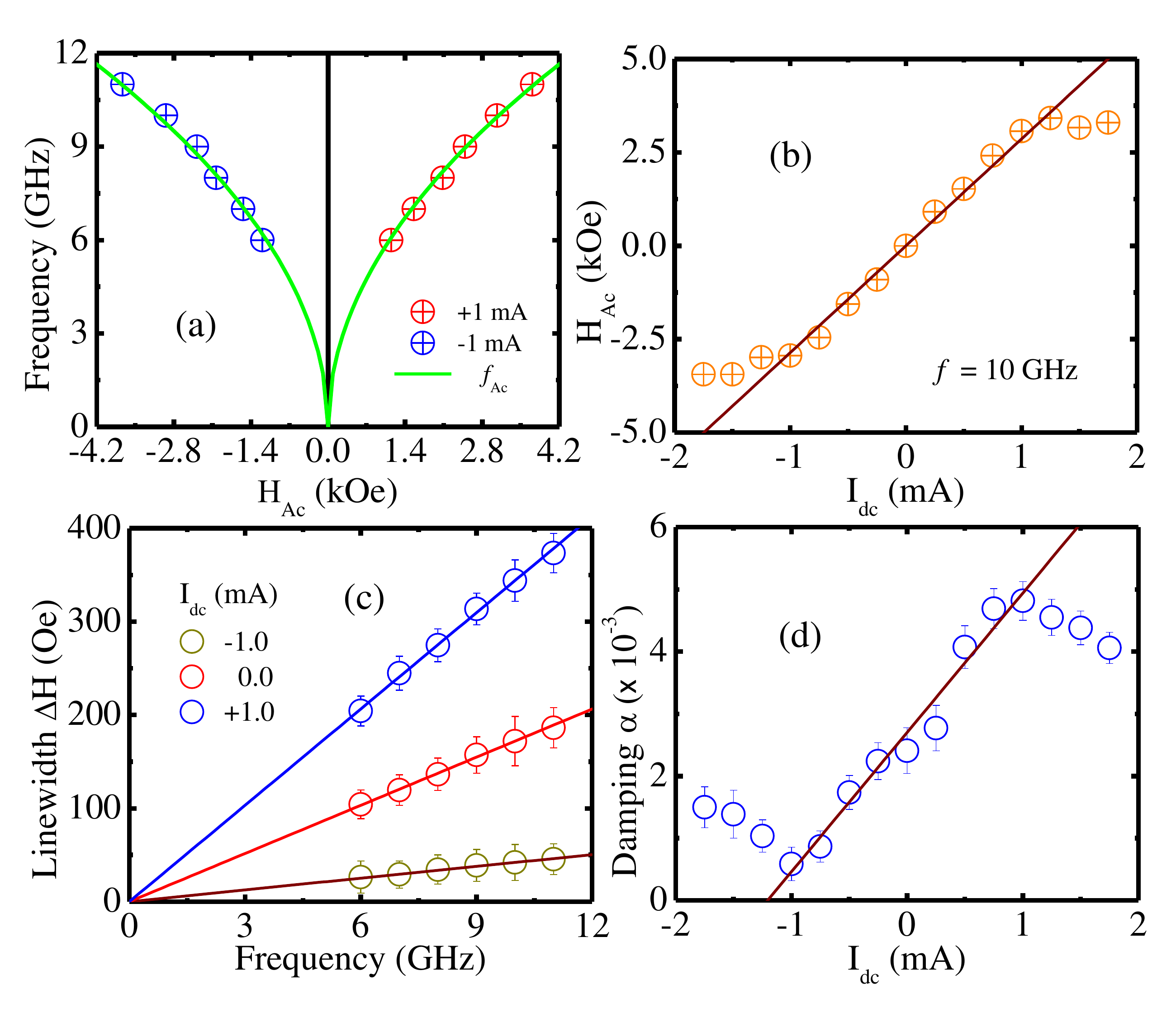}
		\caption{\label{arttype}(Color online) \textbf{(a)} Ferromagnetic resonance (FMR) frequency as a function of the spin accumulation field H$_{Ac}$. The fits are performed with the Kittel equation. \textbf{(b)} Spin accumulation field as a function of the electric current to the frequency of 10 GHz. \textbf{(c)} Shows the linewidth change as a function of the FMR frequency to three values of electric current -1, 0, +1 mA. \textbf{(d)} Damping change as a function of electric current.}
		\label{felicidade}
	\end{center}
\end{figure}

In the \textbf{Fig. 4 (d)}, we show the change of the damping of the IrMn$_{3}$/Pt bilayer as a function of the electric current. Damping starts to increase for electric currents below -1 mA, as well as decreases for electric currents above +1 mA. This behavior agrees with the data of the \textbf{Fig. 4 (b)}, and is due to the spins-in and spins-out configurations in the IrMn$_{3}$/Pt bilayer that loses their dynamic characteristics due to the saturation of the spin accumulation field and the fact that the interface presents a local temperature change [12, 19]. In summary, we experimentally observed a magnetic interfacial effect in IrMn$_{3}$/Pt bilayer, and we manipulated it through the FMR technique without and with electric current.

\section*{Acknowledgements}

This research was supported by Conselho Nacional de Desenvolvimento Cien-tífico e Tecnológico (CNPq), Coordenação de Aperfeiçoamento de Pessoal de Nível Superior (CAPES), Financiadora de Estudos e Projetos (FINEP), and Fundação de Amparo à Ciência e Tecnologia do Estado de Pernambuco (FACEPE).

\bibliographystyle{MiKTeX}

\end{document}